# Transformation and reconstruction towards two-dimensional atomic laminates

Zhiguo Du[1,3], Zongju Cheng[1,3], Qi Zhao[1], Haiyang Wang[1], Qi Zhu[1], Hao Chen[1], Xiao Chen[2], Bin Li[1] and Shubin Yang[1]*

[1]School of Materials Science and Engineering, Beihang University, Beijing, China.
[2]Beijing Key Laboratory of Green Chemical Reaction Engineering and Technology, Department of Chemical Engineering, Tsinghua University, Beijing, China.
[3]These authors contributed equally.
*E-mail: yangshubin@buaa.edu.cn

**Two-dimensional (2D) nanomaterials derived from non-van der Waals solids are promising due to their fantastic physical and chemical properties, but it remains challenging to obtain 2D atomic laminates with high stability owing to the strong intrinsic covalent/metallic bonds and highly exposed active surface. Here, we report a versatile and scalable protocol to produce 2D atomic laminates, based on an unexpected topological transformation of MAX phases under hydrogen chloride gas, and/or subsequent reconstruction under some programmed gases/vapors. In contrast to the known approaches with liquid or molten medium, our method involves in a gas-phase reaction with fast thermodynamics for A layers and positive Gibbs free energies for MX slabs. Remarkably, through subsequent reconstruction in some active gases/vapors ($O_2$, $H_2S$, P, $CH_4$, Al and Sn metal vapors), a big family of 2D atomic laminates with elusive configurations as well as high chemical/thermal stabilities and tunable electrical properties (from metallic to semiconductor-like behaviors) are achieved. Moreover, the resultant 2D atomic laminates can be facilely scaled up to 10 kilograms. We believe that the 2D atomic laminates would have broad applications in catalysis, energy storage, electromagnetic shielding interface and microwave absorption.**

Two-dimensional (2D) nanomaterials derived from van der Waals (vdW) and non-vdW solids have gained great interests owing to their unique physical and chemical properties[1-4]. The atomic layers from vdW solids such as graphene and transition metal disulfides without dangling bonds can be facilely produced based on their vdW counterparts[5,6]. In contrast, producing atomic layers from non-vdW solids are difficult owing to their strong chemical bonds in bulk[2,7,8]. In a near decade, non-vdW solids such as MAX phases open up a new space to produce 2D atomic layers or laminates including transition metal carbides, nitrides and carbonitrides (MXenes) with a hexagonal close-packed crystal structure in a *P*63/*mmc* space group, in which the honeycomb-like transition metal layers are interleaved by carbon and/or nitrogen layers in the octahedral sites[9-11]. Owing to the highly exposed transition metal planes and unsaturated coordination of the surface transition metals, MXenes are commonly covered by terminations (e.g., = O, -OH, -F and -Cl) centered above the outer M layers (face-centered cubic site) or on the top of X atoms (hexagonal-close-packed site)[4,12]. Moreover, it has been demonstrated that the intrinsic features of MXenes are strongly reliant on their atomic configurations such as ordered or solid solution form and tunable surface terminations, enabling fantastic properties such as tailored electrical properties (metallic, semiconductive and even superconductive), good mechanical strength, hydrophilicity and magnetism[13-16]. As a result, MXenes have shown promising applications in electronics, catalysis, energy storage and electromagnetic interface shielding[4,17,18].

Initially, MXenes, one of 2D atomic laminates, were synthesized by using an aqueous solution of HF acid to dissolve A species from MAX phases, associated with soluble A-containing products[19]. The liquid-etching approach was further extended to utilize fluoride salt (e.g., LiF) and acid (e.g., hydrochloric acid) mixture[20], $NH_4HF_2$ or tetramethylammonium hydroxide (TMAOH) solution[21,22], generating a family of F-/OH-terminated MXenes ($M_{n+1}X_nT_x$, M is an early transition metal, X is C and/or N, $T_x$ is the surface termination). Subsequently, molten salt approach was explored by treatment of MAX phases based on the redox potential coupling of A species with molten salts ($ZnCl_2$, $CuCl_2$ or $CdBr_2$)[23-25]. The A layers with lower redox potentials can be extracted from MAX phases by the cations of molten salts with higher redox potentials, affording Cl- and Br-terminated MXenes after an inevitable tedious washing treatment[24,25]. Such F-free MXenes allow to be further tailored with -S, -Se and -Te functional groups through post substitution reactions in an organic system, resulting in fascinating electronic properties such as superconductivity[25]. However, producing MXenes and the beyond atomic laminates with high stabilities especially in a large scale remains a big challenge, greatly hampering their practical applications.

**Transformation towards 2D atomic laminates under HCl gas**

Here, we demonstrate a versatile and scalable approach to produce 2D atomic laminates based on topological transformation of MAX phases under hydrogen chloride (HCl) gas, and/or subsequent reconstruction under some gases/vapors. The key of our approach is utilizing HCl gas, whose molecules enable to directly and rapidly react with A species in MAX phases at high temperatures (> 873 K), associated with the main products of 2D atomic laminates and gaseous by-products of ACl$_y$ (M-A-X + HCl (g) → M-X-Cl + ACl$_y$ (g), **Fig. 1**a and Supplementary Figs. 1-3). Such topological transformation is mainly determinate by the following three factors: 1) active A (such as Al and Si) species in MAX phases, which enable to react with HCl gas molecules owing to their negative Gibbs free energies of < -51.8 kJ mol$^{-1}$ (Fig. 1d and Supplementary Fig. 4) in a wide temperature range from 800 to 1500 K; 2) gaseous by-products ACl$_y$ with high vapor pressures (Fig. 1c), which would be immediately run off from the reaction system, allowing a continuous reaction inside the bulk MAX phases; 3) high chemical and thermal stabilities of MX slabs in HCl gas, which enable the existence of the M-X-Cl configurations including Ti-N-Cl, Ti-C-Cl, Nb-C-Cl, Ta-C-Cl, Zr-C-Cl and Hf-C-Cl due to the positive Gibbs free energies of the reactions between MX slabs and HCl (Fig. 1e); otherwise, the main products would be transition metal chlorides (M-Cl) such as in the cases of Cr and V-based MAX phases (Supplementary Fig. 5). Interestingly, the as-prepared 2D atomic laminates can be further transformed or reconstructed under some programmed gases/vapors including O$_2$, H$_2$S, P, CH$_4$, Al and Sn metal vapors. In this principle, 79 atomic laminates including 2D transition metal carbides, carbonitrides, nitrides with elusive configurations, and even 2D MAX phases (MAXenes) are produced (Supplementary Tables 1 and 2). More importantly, it is easy to scale up to 10 kilograms in our lab (Fig. 1b and Supplementary Fig. 1b). Such high throughput production would greatly promote the practical applications of 2D atomic laminates.

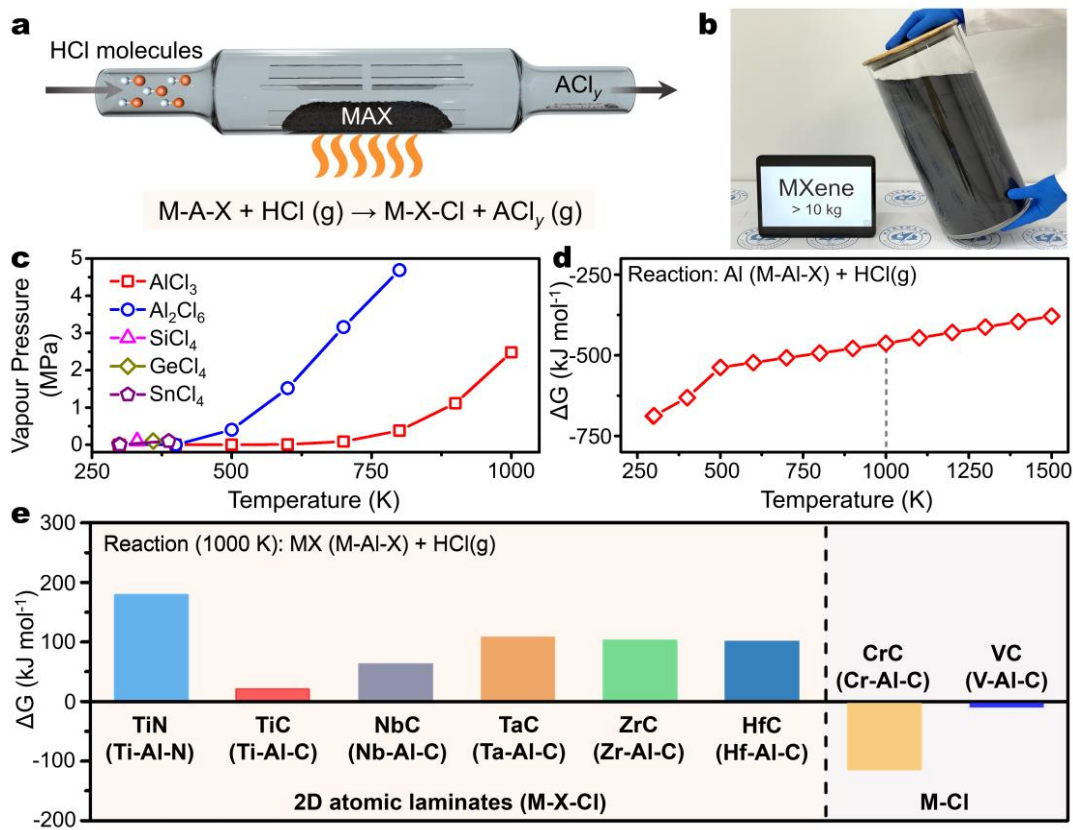

**Fig. 1 | Topological transformation of MAX phases to produce 2D atomic laminates under hydrogen chloride gas. a**, MAX phases are progressively transformed to 2D atomic laminates via the reaction (M-A-X + HCl (g) → M-X-Cl + ACl$_y$ (g)) at high temperatures (873-1073 K), associated with gaseous by-products of ACl$_y$. **b**, Scalable production of > 10 kilograms of 2D atomic laminate (MXene) produced through the topological transformation. **c**, Temperature-vapor pressure relationship of by-product ACl$_y$ derived from MAX phases with Al, Si, Ge or Sn layers, exhibiting Al-based by-product has the highest vapor pressure at a high temperature above 500 K. **d**, Gibbs free energy-temperature plot of the reaction between Al (M-Al-C) layer with HCl gas, disclosing negative Gibbs free energies. **e**, Gibbs free energies of the reactions of MX (M-Al-X) with HCl gas at 1000 K, showing positive Gibbs free energies for Ti-, Nb-, Ta-, Zr- and Hf-based MAX phases and negative Gibbs free energies for Cr- and V-based MAX phases.

As a proof of our concept, we initially chose Ti$_4$AlN$_3$ MAX phase as a precursor to react with HCl gas at 923 K for 25 min to produce nitride 2D atomic laminate (**Fig 2**a; see Methods), since it is relatively difficult to be synthesized by the known approaches. As shown in the scanning electron microscope (SEM) and transmission electron microscopy (TEM) images (Fig. 2b, Supplementary Figs. 2 and 3), accordion-like product is visible, similar to the reported MXenes via liquid-etching approach[26]. This delaminated structure should be originated from the rapid removing of gaseous by-product $AlCl_3$ with high vapor pressures during the topological transformation process (Supplementary Fig. 6). X-ray diffraction (XRD) patterns of the main product exhibit a characteristic peak at 4.7° (Fig. 2c), indexed to the (002) peak of MXene $Ti_4N_3T_x$, which is lower than that of MAX $Ti_4AlN_3$ (7.5°), attributed to the expanded interlayer spacing by extracting Al species. Moreover, the (002) and (004) peaks of $Ti_4AlN_3$ become invisible, demonstrating the complete removal of Al layers from bulk MAX phase. Notably, the complete reaction time is only 25 min, substantially lower than those via liquid-etching (24-48 h) and molten salt (5-24 h) approaches[20,24,26]. Atomic-resolution cross-sectional scanning transmission electron microscope (STEM) images (Fig. 2d and Supplementary Fig. 7) reveal that, in a vertical line direction of a slab, there are four bright atoms in between two legible atoms that should be the chlorine termination. Notably, their interlayer spacings are expanded to 1.31-1.91 nm from previous 1.18 nm ($Ti_4AlN_3$), well consistent with the XRD analysis. Benefited from the loosely packed structure, the resultant nitride could be facilely exfoliated into 2D atomic laminates with a high yield of ~24.5 wt.% via a simple sonication treatment (Fig. 2e and Supplementary Fig. 8). High-resolution transmission electron microscopy (HRTEM) images show well-defined lattice fringes with a distance of 0.27 nm in the delaminated layers (Supplementary Fig. 8e), in precise agreement with the interplanar spacing between (100) facets of $Ti_4N_3Cl_x$. Atomic-resolution STEM images (Fig. 2f and Supplementary Fig. 9) reveal that the transition metal atoms are arranged in a hexagonal symmetry, demonstrating a single crystalline nature of the exfoliated $Ti_4N_3Cl_x$ laminate (Inset in Fig. 2f). The STEM and corresponding elemental mapping images (Supplementary Figs. 8f-i) display the uniform distribution of Ti, N and Cl species on these atomic laminates. The Ti-Cl signals are clearly shown in the X-ray absorption spectroscopy (XAS) and X-ray photoelectron spectroscopy (XPS) analyses (Supplementary Figs. 10 and 11)[23,25]. The content of Cl in the product was evaluated to be 10.4 wt.% from elemental analysis. Based on our topological transformation strategy, a large number of 2D atomic laminates including $Ti_3C_2Cl_x$, $Ti_2CCl_x$, $Nb_2CCl_x$, $Nb_4C_3Cl_x$, $Ta_2CCl_x$, $Ta_4C_3Cl_x$ and $Ti_2NCl_x$, $TiNbCCl_x$ and $Ti_3CNCl_x$ were achieved through the reaction between their counterparts (MAX phases) and HCl gas (Supplementary Figs. 12-23 and Supplementary Table 1). Moreover, the reaction time could be further decreased to 10 min in the cases of some highly active MAX phases such as $Ti_2AlN$, clearly demonstrating the fast thermodynamics of the transformation reaction in HCl gas.

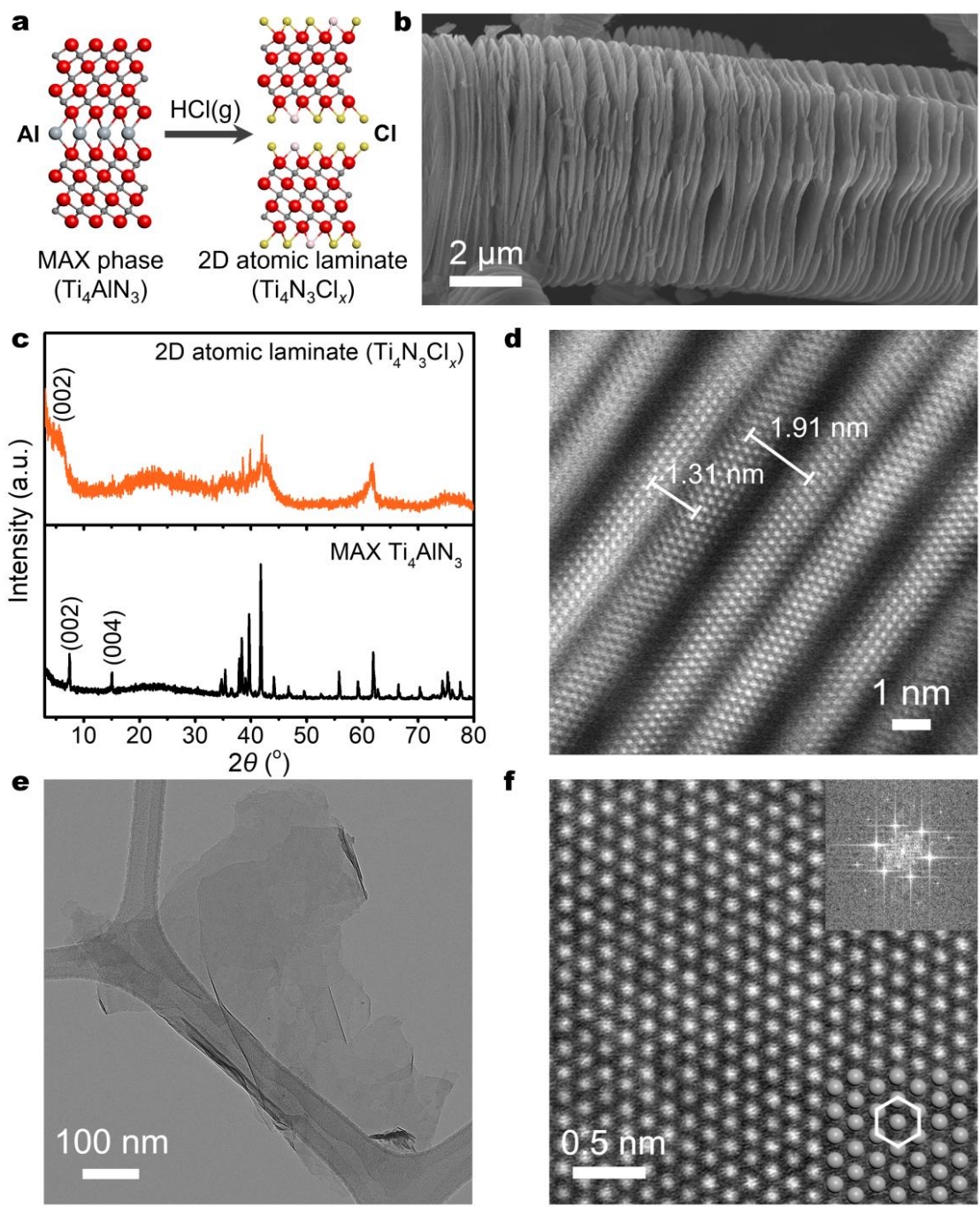

**Fig. 2 | Characterizations of a typical transformed 2D atomic laminate ($Ti_4N_3Cl_x$) under hydrogen chloride gas. a**, Schematic illustration of topological transformation from MAX phase $Ti_4AlN_3$ to 2D atomic laminate $Ti_4N_3Cl_x$ under HCl gas. **b**, SEM image of $Ti_4N_3Cl_x$, showing a typical accordion-like structure. **c**, XRD patterns of MAX $Ti_4AlN_3$ and accordion-like $Ti_4N_3Cl_x$, exhibiting the presence of typical (002) peak for MXene. **d**, Atomic-resolution cross-sectional STEM image of accordion-like $Ti_4N_3Cl_x$, showing the expanded interlayer spacing. **e**, **f**, TEM (**e**) and atomic-resolution STEM (**f**) images of 2D atomic laminate $Ti_4N_3Cl_x$, disclosing the precise atomic structure in a hexagonal system. Insets in (**f**) are the corresponding FFT patterns (top) and atomic configuration (bottom).

**Reconstructed 2D atomic laminates**

Interestingly, such convenient approach allows us to tailor the structural configurations of 2D atomic laminates by further introducing some available gases/vapors including $O_2$, $H_2S$, Se, Te, P and $CH_4$ into the reaction system (Supplementary Figs. 24 and 25). For clearly showing the configuration change, the $M_{n+1}X_nCl_x$ atomic laminate was presented as M-X-Cl in the context (**Fig. 3**a). As introducing chalcogen-containing gases such as $O_2$ into the reaction system at a controllable temperature of 523 K, the atomic laminate with M-X-Cl configurations was tailored to M-X-O in a plane as evidenced in the Fourier transform spectra of Ti K-edge extended X-ray absorption fine structure (EXAFS) spectroscopy (Supplementary Fig. 26). As shown in Fig. 3b, a peak at 2.1 Å indexed to the Ti-Cl bond disappears, along with a stronger peak centered at 1.65 Å (Ti-N/Ti-O bonds)[25,27]. Associated with two increased $t_{2g}$ and $e_g$ peaks at 532.5 and 534.5 eV in the O K-edge X-ray absorption near-edge fine structure (XANES) spectra (Supplementary Fig. 27)[28], respectively, the formation of Ti-O bonds in the resultant 2D Ti-N-O atomic laminates is confirmed. Elemental analysis reveals that the content of O species was increased to 14.0 wt.% from 3.7 wt.% after $O_2$ treatment. Strikingly, as introducing $CH_4$ gas into the reaction system, a new configuration of Ti-N-C emerges in the atomic laminates, as displayed in a C K-edge XANES spectra with an identified peak centered at 283.2 eV (Fig. 3c) for the Ti 3d and C 2p orbital interaction[29]. This can be further demonstrated by the high-resolution C 1s XPS spectra (Supplementary Fig. 28c) with a C-Ti signal at 282.2 eV (ref.[23]). Unlike the reported labile atomic layers, these 2D atomic laminates with Ti-N-O and Ti-N-C configurations possess good tolerances to base solution (up to 3 weeks, Supplementary Figs. 29 and 30) and even high temperature (up to 1100 K, Supplementary Fig. 31), showing good chemical and thermal stabilities. Moreover, it allows us to extend the protocol to synthesize a series of 2D atomic laminates with M-X-S, M-X-Se, M-X-Te, M-X-P configurations (Supplementary Figs. 32-35) as well as Ti-N and Ti-C rocksalt structures (Supplementary Figs. 36 and 37).

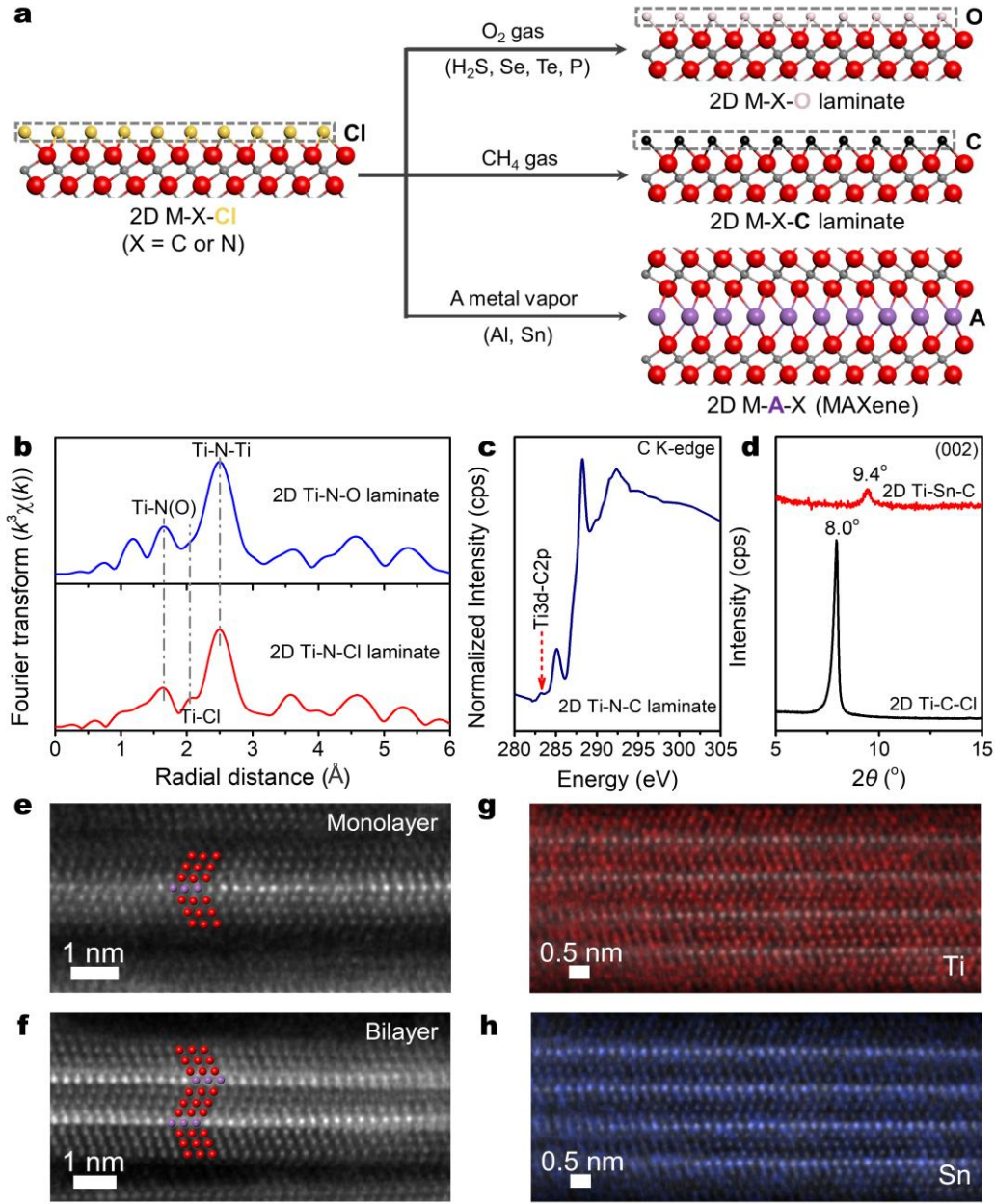

**Fig. 3 | Reconstructed 2D atomic laminates under different gases/vapors. a**, Schematic illustration of the reconstruction of 2D atomic laminates with M-X-O, M-X-C and M-A-X configurations based on 2D M-X-Cl atomic laminates under available gases/vapors ($O_2$, $H_2S$, Se, Te, P, $CH_4$, Al or Sn vapors). **b**, Fourier transform spectra of Ti K-edge EXAFS, exhibiting the disappearance of Ti-Cl peak and the increase of Ti-N/Ti-O peak for 2D Ti-N-O atomic laminates compared to 2D Ti-N-Cl. **c**, C K-edge XANES spectra for 2D Ti-N-C atomic laminate, disclosing the characteristic peak for Ti 3d-C 2p interaction. **d**, Enlarged XRD patterns of 2D Ti-C-Cl and Ti-Sn-C atomic laminates, demonstrating the disappearance of (002) peak for 2D Ti-C-Cl and the presence of broad and weak (002) peak for 2D Ti-Sn-C laminate (MAXene). **e, f**, Atomic-resolution STEM images of monolayer (**e**) and bilayer (**f**) of Ti-Sn-C laminates. **g, h**, Elemental mapping images of 2D Ti-Sn-C atomic laminate, showing the presence of Ti (**g**) and Sn (**h**) species.

Meanwhile, we tried to introduce other vapors such as Sn metal vapor into the reaction system to study if it has a high activity with M-X-Cl atomic laminates like above available gases. Remarkably, after reaction at a high temperature of 1073 K, the (002) peak at 8.0° of 2D Ti-C-Cl disappeared in the XRD patterns (Fig. 3d and Supplementary Fig. 38), instead of a new weak and broad peak appeared at 9.4°, indexed to the (002) facet reported for $Ti_3SnC_2$ MAX phase[30]. The reconstructed atomic laminates can be further evidenced by the atomic-resolution STEM images (Fig. 3e, 3f), where one bright atomic (Sn) layer is precisely occupied into the adjacent MX slabs with a distance of 1.0 nm (Fig. 3g, 3h). Thus, it gives rise to new 2D atomic laminates with monolayer (Fig. 3e), bilayer (Fig. 3f) and quadrilayer (Supplementary Fig. 39b). Such precise structure should be governed by the underlying crystalline symmetry/lattice. Unexpectedly, the exposed surface of the 2D atomic laminates was still terminated with chlorine rather than the metallic Sn atoms (Supplementary Fig. 39c). These results suggest that there should be a zipper effect during the reaction between Sn vapor and 2D M-X-Cl atomic laminates, i.e., the metal Sn atoms only have a function to zipper the adjacent MX slabs with low stereo-hindrance (Supplementary Fig. 39b), rather than substituting the chlorine terminations. This process can be regarded as a precise atomic additive manufacturing in between adjacent atomic layers, possibly enabling to synthesize many elusive 2D atomic laminates with precise structures (such as 2D $Ti_3AlC_2$ atomic laminates by introducing Al metal vapors, Supplementary Figs. 40-44). The structural configurations of atomic laminates would have a large influence on their electrical properties[31]. As introducing metal Sn and Al vapors into the reaction system, the reconstructed 2D atomic laminates have high electrical conductivities up to 26800 and 47600 S m$^{-1}$, respectively, higher than that of the commercial $Ti_3AlC_2$ powder (4390

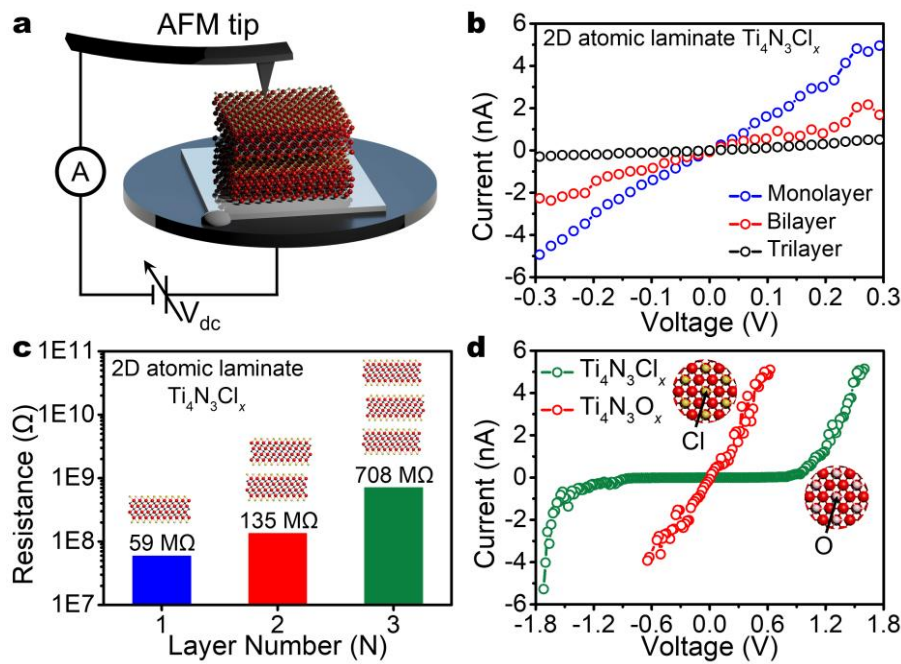

**Fig. 4 | Electrical properties of 2D atomic laminates with different configurations.**
**a**, Schematic of electrical measurements of 2D atomic laminates by PF-TUNA setup in AFM. **b**, Current-voltage curves measured through 2D atomic laminate $Ti_4N_3Cl_x$ with different layer numbers, disclosing a linear relationship between the current and voltage. **c**, Vertical resistances of 2D atomic laminate $Ti_4N_3Cl_x$ with monolayer, bilayer and trilayer. Insets are the corresponding atomic configurations. **d**, Current-voltage curves measured through 2D $Ti_4N_3Cl_x$ and $Ti_4N_3O_x$, showing a typical semiconductor-like relationship for the 2D atomic laminate $Ti_4N_3O_x$. Insets are the atomic models of 2D $Ti_4N_3Cl_x$ and $Ti_4N_3O_x$ from top view.

S m$^{-1}$). These values are much higher than the atomic layers with M-X-Cl (12800 S m$^{-1}$) and M-X-O (160 S m$^{-1}$) configurations. To gain further insight into the contact resistance of the 2D atomic laminates with M-X-Cl and M-X-O configurations (Supplementary Figs. 45 and 46), we conducted a peak force tunneling (PF-TUNA) measurement in atomic force microscopy (AFM) (**Fig. 4a**). Clearly, there are distinct tunneling currents through the Ti-N-Cl atomic laminates in the vertical direction, and the current and voltage are in a linear character (Fig. 4b), consistent with the direct tunneling current in the low bias regime[32]. The vertical resistances of Ti-N-Cl bilayer and trilayer are 135 and 708 MΩ (Fig. 4c), respectively, much higher than the monolayer (59 MΩ), disclosing that the layer-to-layer contact resistance largely increases with increasing the layer numbers[33]. In contrast, atomic laminates with Ti-N-O configuration show a non-linear voltage-current relationship (Fig. 4d and Supplementary Fig. 47), where there is negligible current in the low bias range and the tunneling current nonlinearly increases and finally reaches the limited current as further increasing the voltage bias. There is an approximatively linear relationship between $\ln(I/V_2)$ and $1/V$ (Supplementary Fig. 48), suggesting that the electron transport is dominated by the Fowler-Nordheim (F-N) tunneling, displaying a semiconductor-like behavior, similar to that of 2D transition-metal chalcogenide layers[34,35]. This metallic-semiconductive transition should be originated from the tailored electronic structure of Ti-N-O atomic laminates caused by the surface coverage with O species.

**A convenient and scale-up synthetic route**

Tedious washing procedures are often inevitable for producing MXenes and the beyond atomic laminates for the purpose of removing by-products and residues. Although it can be carried out in a research lab, it represented a major hurdle for industrial scale-up. In contrast, our atomic laminates are directly produced in a gaseous system with the fast removal of gaseous by-product, leaving pure main product. Such simple washing-free processes might prove decisive for the viable commercial scale-up of 2D atomic laminates with low costs. As calculated, the cost of our 2D atomic laminates mainly comes from the commercially available raw materials (MAX phases and HCl gas) as shown in Supplementary Table 3. In the case of our accordion-like MXene ($Ti_3C_2Cl_x$), its cost is only 0.2 $ g$^{-1}$, only 1/100 of the price of commercially available product. Moreover, based on their high exfoliation yields, the cost of monolayer and few-layer MXenes would be less than 2.0 $ g$^{-1}$. It is anticipated that, with decreasing the prices of commercial MAX phases, cost savings should be further achieved, which would greatly promote the applications of 2D atomic laminates in wide areas of catalysis, energy storage, electromagnetic shielding interface and microwave absorption (Supplementary Fig. 49).

**References**

1. Chowdhury, T., Sadler, E. C. & Kempa, T. J. Progress and prospects in tansition-metal dichalcogenide research beyond 2D. *Chem. Rev.* **120**, 12563-12591, (2020).
2. Balan, A. P. *et al.* Exfoliation of a non-van der Waals material from iron ore hematite. *Nat. Nanotech.* **13**, 602-609, (2018).
3. Lin, Z. *et al.* Solution-processable 2D semiconductors for high-performance large-area electronics. *Nature* **562**, 254-258, (2018).
4. Mohammadi, A. V., Rosen, J. & Gogotsi, Y. The world of two-dimensional carbides and nitrides (MXenes). *Science* **372**, eabf1581, (2021).
5. Coleman, J. N. *et al.* Two-dimensional nanosheets produced by liquid exfoliation of layered materials. *Science* **331**, 568-571, (2011).
6. Pinilla, S., Coelho, J., Li, K., Liu, J. & Nicolosi, V. Two-dimensional material inks. *Nat. Rev. Mater.* **7**, 717-735 (2022).
7. Gogotsi, Y. & Anasori, B. The rise of MXenes. *ACS Nano* **13**, 8491-8494, (2019).
8. Sokol, M., Natu, V., Kota, S. & Barsoum, M. W. On the chemical diversity of the MAX phases. *Trend. Chem.* **1**, 210-223, (2019).
9. Anasori, B., Lukatskaya, M. R. & Gogotsi, Y. 2D metal carbides and nitrides (MXenes) for energy storage. *Nat. Rev. Mater.* **2**, 16098, (2017).
10. Li, X. *et al.* MXene chemistry, electrochemistry and energy storage applications. *Nat. Rev. Chem.* **6**, 389-404, (2022).
11. Naguib, M., Barsoum, M. W. & Gogotsi, Y. Ten years of progress in the synthesis and development of MXenes. *Adv. Mater.* **33**, 2103393, (2021).
12. Lim, K. R. G. *et al.* Fundamentals of MXene synthesis. *Nat. Synthesis* **1**, 601-614, (2022).
13. Wyatt, B. C., Rosenkranz, A. & Anasori, B. 2D MXenes: Tunable mechanical and tribological properties. *Adv. Mater.* **33**, 2007973, (2021).
14. Hantanasirisakul, K. *et al.* Evidence of a magnetic transition in atomically thin

Cr$_2$TiC$_2$T$_x$ MXene. *Nanoscale Horiz.* **5**, 1557-1565, (2020).

15. Zhou, J. *et al.* Two-dimensional hydroxyl-functionalized and carbon-deficient scandium carbide, ScC$_x$OH, a direct band gap semiconductor. *ACS Nano* **13**, 1195-1203, (2019).
16. Tao, Q. *et al.* Two-dimensional Mo$_{1.33}$C MXene with divacancy ordering prepared from parent 3D laminate with in-plane chemical ordering. *Nat. Commun.* **8**, 14949, (2017).
17. Verger, L., Natu, V., Carey, M. & Barsoum, M. W. MXenes: An introduction of their synthesis, select properties, and applications. *Trend. Chem.* **1**, 656-669, (2019).
18. Iqbal, A. *et al.* Anomalous absorption of electromagnetic waves by 2D transition metal carbonitride Ti$_3$CNT$_x$ (MXene). *Science* **369**, 446-450, (2020).
19. Naguib, M. *et al.* Two-dimensional nanocrystals produced by exfoliation of Ti$_3$AlC$_2$. *Adv. Mater.* **23**, 4248-4253, (2011).
20. Ghidiu, M., Lukatskaya, M. R., Zhao, M. Q., Gogotsi, Y. & Barsoum, M. W. Conductive two-dimensional titanium carbide 'clay' with high volumetric capacitance. *Nature* **516**, 78-81, (2014).
21. Feng, A. *et al.* Fabrication and thermal stability of NH$_4$HF$_2$-etched Ti$_3$C$_2$ MXene. *Ceram. Int.* **43**, 6322-6328, (2017).
22. Xuan, J. *et al.* Organic-base-driven intercalation and delamination for the production of functionalized titanium carbide nanosheets with superior photothermal therapeutic performance. *Angew. Chem. Int. Ed.* **55**, 14569-14574, (2016).
23. Li, M. *et al.* Element replacement approach by reaction with Lewis acidic molten salts to synthesize nanolaminated MAX phases and MXenes. *J. Am. Chem. Soc.* **141**, 4730-4737, (2019).
24. Li, Y. *et al.* A general Lewis acidic etching route for preparing MXenes with enhanced electrochemical performance in non-aqueous electrolyte. *Nat. Mater.* **19**, 894-899, (2020).
25. Kamysbayev, V. *et al.* Covalent surface modifications and superconductivity of two-dimensional metal carbide MXenes. *Science* **369**, 979-983, (2020).
26. Lukatskaya, M. R. *et al.* Cation intercalation and high volumetric capacitance of two-dimensional titanium carbide. *Science* **341**, 1502-1505, (2013).
27. Djire, A., Zhang, H., Reinhart, B. J., Nwamba, O. C. & Neale, N. R. Mechanisms of hydrogen evolution reaction in two-dimensional nitride MXenes using in situ X-ray absorption spectroelectrochemistry. *ACS Catal.* **11**, 3128-3136, (2021).
28. Kao, L. C. *et al.* Trace key mechanistic features of the arsenite sequestration reaction with nanoscale zerovalent iron. *J. Am. Chem. Soc.* **143**, 16538-16548, (2021).
29. Minasian, S. G. *et al.* Carbon K-edge X-ray absorption spectroscopy and time-dependent density functional theory examination of metal-carbon bonding in metallocene dichlorides. *J. Am. Chem. Soc.* **135**, 14731-14740, (2013).
30. Dubois, S., Cabioc'h, T., Chartier, P., Gauthier, V. & Jaouen, M. A new ternary nanolaminate carbide: Ti$_3$SnC$_2$. *J. Am. Ceram. Soc.* **90**, 2642-2644, (2007).
31. Hart, J. L. *et al.* Control of MXenes' electronic properties through termination and

intercalation. *Nat. Commun.* **10**, 522 (2019).
32. Liao, M. *et al.* Twist angle-dependent conductivities across $MoS_2$/graphene heterojunctions. *Nat. Commun.* **9**, 4068, (2018).
33. Fu, D. *et al.* Mechanically modulated tunneling resistance in monolayer $MoS_2$. *Appl. Phys. Lett.* **103**, 183105, (2013).
34. Li, Y., Xu, C. Y. & Zhen, L. Surface potential and interlayer screening effects of few-layer $MoS_2$ nanoflakes. *Appl. Phys. Lett.* **102**, 143110, (2013).
35. Son, Y. *et al.* Layer number dependence of $MoS_2$ photoconductivity using photocurrent spectral atomic force microscopic imaging. *ACS Nano* **9**, 2843-2855, (2015).

**Acknowledgments:** This work was financially supported by National Natural Science Foundation of China (Grant Nos. 52125207 and 52202204), Beijing Natural Science Foundation (Grant No. JQ20011) and National Postdoctoral Program for Innovative Talents (Grant No. BX20200027). We thank the support from Beijing Synchrotron Radiation Facility and National Synchrotron Radiation Laboratory (NSRL, Hefei, China).

**Author contributions:** S.Y. supervised the project. Z.D. and Z.C. designed and carried out all of the experiments. H.W., Q.Z. performed the SEM, TEM, XAS and XRD measurements. B.L. and Q.Z. carried out the XPS and Raman analysis. X.C. performed the STEM measurements. Z.D. carried out the AFM measurements. H.C. carried out the electromagnetic measurement. All authors discussed the results and assisted during manuscript preparation.

**Competing interests:** Authors declare no competing interests.

**Additional information**

**Supplementary information** is available for this paper.

**Reprints and permissions information** is available at www.nature.com/reprints.

**Correspondence and requests for materials** should be addressed to S.Y.

**Methods**

Synthesis of MAX phases

MAX phases were mainly synthesized by a traditional powder metallurgy method[36,37]. The details have been listed in Materials and Methods in Supplementary Information.

Synthesis of 2D atomic laminates with M-X-Cl configuration

Atomic laminates with M-X-Cl configuration ($M_{n+1}X_nCl_x$ MXenes) were synthesized by topological transformation of MAX phases under HCl gas at high temperatures from 900 to 1200 K for 10-30 min. Taking 2D Ti-N-Cl atomic laminates as an example, 200 mg of $Ti_4AlN_3$ phases was reacted at 923 K for 25 min under an HCl/Ar gas at a flow rate of 100/50 sccm.

Reconstruction of 2D atomic laminates with tunable configurations

2D atomic laminates with tunable configurations were synthesized by the reaction of 2D Ti-N-Cl with some gases/vapors including $O_2$, $H_2S$, Se, Te, P and $CH_4$. In the case of 2D Ti-N-O atomic laminates, 300 mg of MXene $Ti_4N_3Cl_x$ was placed in a muffle furnace and heated to 523 K for 3 h with a rate of 3 K $min^{-1}$. In the case of 2D Ti-N-C atomic laminates, 300 mg of $Ti_4N_3Cl_x$ was heated at 20 K $min^{-1}$ under Ar flow of 50 sccm, and 10 sccm of $CH_4$ mixture was injected when the temperature reached 923 K. MXene $Ti_4N_3Cl_x$ was held at this temperature for 10 min, producing 2D atomic laminate with Ti-N-C configuration. The detailed conditions for synthesis of 2D atomic laminates with other configurations are described in Materials and Methods in Supplementary Information.

Reconstructed 2D atomic laminates under metal vapors (MAXenes)

2D MAX phases (MAXenes) were synthesized by the reaction of 2D M-X-Cl atomic laminates with A metal vapors such as Al and Sn at a high temperature of 973 and 1073 K. Taking 2D MAX phase $Ti_3SnC_2$ as an example, 100 mg of MXene $Ti_3C_2Cl_x$ was uniformly dispersed on the surface of a piece of carbon paper and placed on the top of a porcelain boat with 500 mg Sn powder inside. The porcelain boat was heated to 1073 K with a rate of 10 K min$^{-1}$ under Ar flow of 50 sccm and held at that temperature for 12 h, producing 2D $Ti_3SnC_2$ MAX phase. Similarly, 2D $Ti_3AlC_2$ MAX phase was synthesized by using 100 mg of Al powder.

Vapor pressure calculations

The vapor pressures of $ACl_y$ intermediates were calculated based on two-phase equilibrium reaction:

$ACl_y$ (solid, liquid) = $ACl_y$ (gas)

Based on the log $K_f$ values, the vapor pressure of $ACl_y$ at a given temperature of T was calculated as following:

log $K$ = log $K_f(ACl_y$ (gas)) − log $K_f(ACl_y$ (solid, liquid)) = log $[p(ACl_y$ (gas))/$a(ACl_y$ (solid, liquid))]

where $p$ is the vapor pressure, $a$ is the activity of $ACl_y$ in the solid or liquid phase, $K_f$ is the equilibrium constant. The referred log $K_f$ values[38] are listed in the Supplementary Table 4. Specifically, in the case of the intermediate in condensed phase, the value of $a$ is 1.

The evaporation equation can be simplified:

log $K_f(ACl_y$ (gas)) − log $K_f(ACl_y$ (solid, liquid)) = log $p(ACl_y$ (gas))

and

$p(ACl_y \text{ (gas)}) = 10^{\log K_f(ACl_y \text{ (gas)}) - \log K_f(ACl_y \text{ (solid, liquid)})}$

where $p$ is in units of bar (1 bar = $10^5$ Pa).

Gibbs free energy ($\Delta G$) calculations

At a specific temperature of T, the Gibbs free energy ($\Delta G$) of a given chemical reaction was calculated to be the algebraic sum of Gibbs free energy of each component times its stoichiometric number. Specifically, the $\Delta G$ is equal to the sum of Gibbs free energy for all the products minus the sum of Gibbs free energy for all the reactants. The reaction of A species of MAX phases with HCl gas is calculated based on the following equation:

A + z HCl (gas) = $ACl_y$ + z/2 $H_2$ (gas)

In the case of Al species:

Al + 3 HCl (gas) = $AlCl_3$ + 3/2 $H_2$ (gas)

The $\Delta G$ of the above reaction can be calculated according to the following formula:

$\Delta G = [1.5\Delta G (H_2 \text{ (gas)}) + \Delta G (AlCl_3)] - [3 \Delta G (HCl \text{ (gas)}) + \Delta G (Al)]$

where $\Delta G$ is the Gibbs energy at a given temperature of T. The $\Delta G$ values are taken from literature[38] and listed in Supplementary Table 5. If the $\Delta G$ value of the reaction is negative, it should enable to proceed at this temperature.

The stability of MX slab of MAX phases at 1000 K is evaluated based on the reaction of binary phase MX with HCl gas in the following:

MX + z HCl (gas) = $MCl_y$ + X + z/2 $H_2$ (gas)

In the case of MXene $Ti_4N_3$:

TiN + 2 HCl (gas) = $TiCl_2$ + 1/2 $N_2$ (gas) + $H_2$ (gas)

The ΔG of the above reaction at 1000 K can be calculated according to the following formula:

$$\Delta G = [\Delta G\ (TiCl_2) + \Delta G\ (C) + 1/2\ \Delta G\ (N_2\ (gas))] – [2\ \Delta G\ (HCl\ (gas)) + \Delta G\ (TiN)]$$

where ΔG is the Gibbs energy for each component at 1000 K. The ΔG values are taken from literature and listed in Supplementary Tables 6 and 7. If the ΔG value is positive, indicating that the reaction will not occur at 1000 K, demonstrating the high chemical and thermal stability of the resultant 2D atomic laminates.

**Data availability**

The data that support the findings of this study are available from the corresponding authors on reasonable request.

36. Urbankowski, P. *et al.* Synthesis of two-dimensional titanium nitride $Ti_4N_3$ (MXene). *Nanoscale* **8**, 11385-11391, (2016).
37. Naguib, M. *et al.* New Two-Dimensional Niobium and Vanadium Carbides as Promising Materials for Li-Ion Batteries. *J. Am. Chem. Soc.* **135**, 15966-15969, (2013).
38. Barin, I. *Thermochemical Data of Pure Substances* 3rd edn (VCH, 1995).